# Nonlinear optical responses of organic based indole derivative: an experimental and computational study

Meenakshi Rana[a], Papia Chowdhury[b*]

[a]*Department of Physics, School of Sciences, Uttarakhand Open University, Haldwani, 263139, Nainital, India*
[b]*Department of Physics and Materials Science & Engineering, Jaypee Institute of Information Technology, Noida 201307, Uttar Pradesh, India*

**Abstract**

The nonlinear optical (NLO) properties of the Indole-7-carboxyldehyde (I7C) have been investigated by computational and experimental (UV-VIS, Raman) data analysis. Mulliken charge analysis, molecular electrostatic potential, and UV-VIS absorption and vibrational Raman studies have been used to analyze the intra-molecular charge transfer occurrence in the probe system. Observed high value of dipole moment, linear polarizability and first order hyperpolarizability values suggest that the indole derivative may indeed have possibility to show good NLO behaviour.

## 1. Introduction

Nonlinear optical active (NLO) materials have attracted considerable attention in many applications such as integrated photonics, optical rectification, sum and difference frequency generation (SFG, DFG), second harmonic generation (SHG), frequency mixing terahertz (THz) wave generation, telecommunication [1-3] etc. NLO materials are very unique in nature. These materials can produce a modified field after interacting with an externally applied field [4]. Applications of NLO materials in industry have been observed from 1961 when SHG was first observed in single crystal quartz by Franken and co-workers. Since then many research works have been dedicated to the synthesize and designing of organic/inorganic efficient NLO materials with higher polarizability ($\alpha$), first order hyperpolarizability ($\beta$). The presence of π conjugation in organic systems [5,6] which include alternate single and double bond for the molecular structure formation are the best candidate to show the NLO activity. It is well known that a material exhibiting NLO activity must be noncentrosymmetric (NCS). Indole-7-carboxyaldehyde (I7C) is the indole substituted doped π conjugated noncentrosymmetric organic systems.

Nowadays Indole and their derivatives are very attractive topic of research. They are similar with tryptophan residue, which is a part of protein as RNA and DNA [7]. I7C is one of such indole derivative that have N–H⋯O=C type H-bonding. I7C possesses a number of interesting chemical and physical properties that validate its tremendous applications in many fields such as; vitamin supplements, over-the-counter drugs, flavour enhancers, and perfumery [8]. The application of indole and their substituted derivatives have been extended to reduce breast or prostate cancer. In the present work, we have reported the NLO properties of I7C. The structural geometry, charge (Mulliken) analysis, frontier molecular orbitals (FMOs) and molecular electrostatic potential (MEP) indicate the delocalization of charges over the donor acceptor region. The vibrational Raman analysis confirms the charge transfer interaction between donor and acceptor groups, and that in turn validates the presence of the larger dipole moment ($\mu$), polarizability and hyperpolarizabilities ($\alpha$ and $\beta$) in I7C.

* Corresponding author.
*E-mail address:* papia.chowdhury@jiit.ac.in

## 2. Experimental Section
### 2.1 Materials and Methods

I7C was purchased from Aldrich Chemical, USA and solvent teterahydrofuran (THF) was purchased from Fluka. UV-Vis absorption spectra were recorded at 300 K with a Perkin-Elmer absorption spectrophotometer (model Lamda-35, source: tungsten iodide and deuterium) in the spectral range 200−1100 nm with a varying slit width (0.5 and 1.0 nm). To record Raman spectra, a 532 nm excitation diode pumped frequency doubled Nd:YAG solid state laser (model GDLM-5015 L, 8mW) have been used.

*2.2 Theoretical Methods*

All the theoretical calculations, including the optimization of ground and excited state geometries were carried out using the Gaussian 09 program [9]. The geometry of I7C are fully optimized, with density functional theory (DFT) using Becke3-Lee-Yang-Parr (B3) exchange functional combined with the (LYP) [37] correlation functional with the standard 6-311G basis set. Excited states of molecules were performed using DFT and TDDFT. All the frequency calculations were carried out using the optimized structures. To visualize the vibrational normal modes, Gauss View 5 molecular visualization program was used [10]. The computed Raman activities are converted to corresponding relative Raman intensities using the following relationship derived from the intensity of Raman scattering.

$$I_i = \frac{f(\upsilon_0 - \upsilon_i)^4 S_i}{\upsilon_i[1 - \exp(-\frac{hc\upsilon_i}{KT})]}, \quad (1)$$

where $\upsilon_0 = 9398.5$ cm$^{-1}$ is the laser exciting wavenumber, $\upsilon_i$ is the vibrational wavenumber of the $i^{th}$ normal mode (cm$^{-1}$), $S_i$ is the Raman scattering activity of normal mode $\upsilon_i$, and $f$ is a constant (in our calculation $f$ is $10^{-12}$), which is suitably chosen common normalized factor for all peak intensities; $h$, $k$, $c$, and $T$ are the Planck constant, the Boltzmann constant, the speed of light and the temperature in Kelvin, respectively.

The FMOs and first order hyperpolarizability are also calculated. Further, molecular electrostatic potential (MEP) surface mapped with electrostatic potential surface and atomic charges (Mulliken) were derived by using optimized structure.

**3. Results and Discussion**

*3.1. Structure*

I7C is a five membered indole substituted probe system having pyrrole (N−H) and carbonyl (C=O) groups. The optimized structure of I7C shown in figure 1. Cis (*C*) and trans (*T*) are the two conformeric forms of the I7C having different geometrical parameters (Figure 1). Out of these two forms, *C* form is more preferable than *T* form [11].

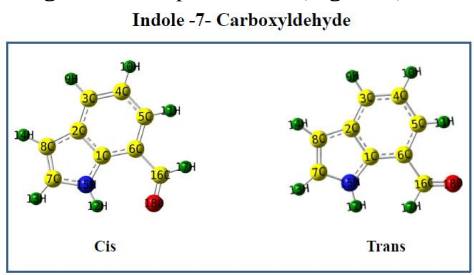

**Figure 1. Optimized cis and trans structure of I7C.**

**Table 1. Energy and dipole moment of I7C conformers.**

| Parameter | Cis (*C*) | Trans (*T*) |
|---|---|---|
| **Energy (a.u.)** | -13041.88 | -12987.18 |
| **Dipole moment (Debye)** | 1.89 | 3.0437 |

The reason of the stability of *C* form over *T* is the strong I$_{ra}$HB between carbonyl and pyrrole groups. Energetically also the cis form is more stable (Table 1). Therefore, in the present manuscript we only consider the *C* form of I7C.

*3.2. Charge Analysis*

I7C is NCS with C$_1$ point group symmetry. Graphical representation of Mulliken charge analysis is shown in Figure 2. The results of charge analysis predicts that all the positive charge spread over hydrogen atoms (H) and negative charge over electronegative nitrogen atoms (N) and oxygen (O). Charge analysis predicts that H$_{12}$ atom of the amine group, shows the maximum positive charges as 0.400 e in H atom. However, O$_{18}$ atom of the caronyl group reveals negative charge of -0.459 e. This causes a huge charge variation betwnn -NH group and acceptor C=O group which initiates the possibility of charge transfer (CT) in I7C.

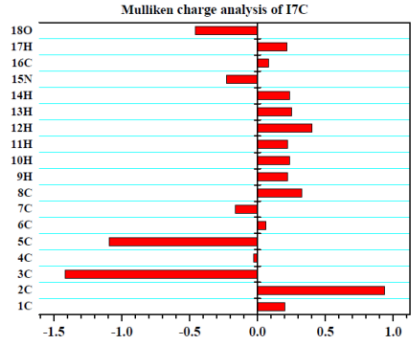

**Figure 2. Mulliken charge plots of I7C.**

*3.3. Nonlinear Optical Analysis*

To study the NLO activeness of I7C, $\alpha$ and anisotropy of polarizability ($\Delta \alpha$) were calculated by using equation (2) and (3):

$$\alpha = \frac{(\alpha_{XX} + \alpha_{YY} + \alpha_{ZZ})}{3} \quad (2)$$

$$\Delta\alpha = 2^{-\frac{1}{2}}[(\alpha_{XX} - \alpha_{YY})^2 + (\alpha_{YY} - \alpha_{ZZ})^2 + (\alpha_{ZZ} - \alpha_{XX})^2 + 6\alpha_{XX}^2]^{\frac{1}{2}} \quad (3)$$

Using finite field theory approach, the value of $\beta$, is computed by using different components of $\beta$ as:

$$\beta_i = \beta_{iii} + \frac{1}{3}\sum(\beta_{iij} + \beta_{jij} + \beta_{jji}), (i \neq j) \quad (4)$$

Using X, Y and Z components of $\beta$, the magnitude of $\beta$ can be calculated as:

$$\beta = (\beta_X^2 + \beta_Y^2 + \beta_Z^2)^{\frac{1}{2}} \quad (5)$$

The final equation for calculating the magnitude of $\beta$ can be written as:

$$\beta = [(\beta_{XXX} + \beta_{XYY} + \beta_{XZZ})^2 + (\beta_{YZZ} + \beta_{YYY} + \beta_{YXX})^2 + (\beta_{ZXX} + \beta_{ZYY} + \beta_{ZZZ})^2]^{\frac{1}{2}} \quad (6)$$

where $\beta_{XXX}$, $\beta_{YYY}$ and $\beta_{ZZZ}$ are the tensor components of hyperpolarizability (Table 2).

**Table 2. Dipole moment, Polarizabilities and First order hyperpolarizabilities of I7C.**

| Dipole moment components | I7C | Polarizabilities components | I7C | First order hyperpolarizabilities components | I7C |
|---|---|---|---|---|---|
| $\mu_X$ | -1.61 | $\alpha_{XX}$ | 143.80 | $\beta_{XXX}$ | 443.11 |
| $\mu_Y$ | 0.97 | $\alpha_{XY}$ | 2.311 | $\beta_{XXY}$ | -90.25 |
| $\mu_Z$ | -0.002 | $\alpha_{YY}$ | 147.06 | $\beta_{XYY}$ | 95.10 |
| $\mu_{total}$ | 1.88 | $\alpha_{XZ}$ | -0.00 | $\beta_{YYY}$ | 0.214 |
| | | $\alpha_{YZ}$ | 0.003 | $\beta_{XXZ}$ | -0.117 |
| | | $\alpha_{ZZ}$ | 60.63 | $\beta_{XYZ}$ | 0.057 |
| | | $\alpha$ | 17.36X10$^{-24}$ esu | $\beta_{YYZ}$ | 0.032 |
| | | $\Delta \alpha$ | 38.83X10$^{-24}$ esu | $\beta_{XZZ}$ | -73.05 |
| | | | | $\beta_{YZZ}$ | 34.69 |
| | | | | $\beta_{ZZZ}$ | -0.049 |
| | | | | $\beta$ | 3.96x10$^{-30}$ esu |

To compute the NLO properties of I7C, we have used equation 2 to 6. The value of $\mu$, $\alpha$, and $\beta$ are obtained as 1.88 Debye, 17.36x10$^{-24}$ esu and 3.96x10$^{-30}$ esu, respectively (Table 2). For I7C, the computed values of $\alpha$ and $\beta$ are almost 3 and 6 times greater than in case of available NLO active inorganic material such as urea. High value of α and β obtained from the computed results confirm the good NLO activeness of the present organic system.

*3.4. Vibrational Analysis*

Raman modes are associated with the NLO activity, since polarizability is proportional to Raman intensity. We have computed Raman intensity for different nonlinear vibrations. Linear Raman spectroscopy is very useful technique for predicting different NLO coefficients like: α, β for the probe system where stronger Raman signal means higher Raman activity and so the higher polarizability. We have added this information in the revised manuscript. High Raman intensity of I7C also show the NLO activeness of the probe system. Table. 3 and Figure 3 shows the experimental and computed Raman frequencies of I7C. Analyses of some selected vibrational modes are as follows:

**Table 3: Experimental and computed Raman data for I7C.**

| S. No | Modes | Theoretical (ν) cm$^{-1}$ | Theoretical Raman intensity | Experimental Raman (cm$^{-1}$) |
|---|---|---|---|---|
| 1. | ν$_{CC}$ δ$_{NH}$ | 1078.13 | 38.8903 | 1063.10 |
| 2. | ν$_{CO}$ | 1721.60 | 118.9435 | 1721.53 |
| 3. | ν$_{CH}$ | 2921.20 | 169.9468 | 2824.45 |
| 4. | ν$_{CH}$ | 3257.26 | 172.0191 | 3137.06 |
| 5. | ν$_{NH}$ | 3625.88 | 107.0070 | 3244.24 |

Vibrational modes: ν, symmetric stretching; δ, bending in plane.

The C-H stretching vibration (ν$_{CH}$) in a heterocyclic aromatic compounds is mainly observed in the region 3200 – 2800 cm$^{-1}$ [12] and are not affected by any aromatic substitutions. Symmetric stretching (ν$_{NH}$) vibrational modes is occur in the region 3200 - 3800 cm$^{-1}$. A strong ν$_{NH}$ band is observed in Raman at 3244.24 cm$^{-1}$ (Figure 3). Experimentally observed in plane bending δ$_{NH}$ mode matches with the computed Raman mode due to δ$_{NH}$. Theoretical ν$_{CO}$ and experimental mode also shows the closer similarity. Since most of the computational data are observed on a single probe whereas experimental finding are based on more than one probe system. So exact matching between experimental output and computational data is not at all possible. Raman mode of ν$_{CN}$ indicates high Raman intensity (172.01, 169.94) and hence high value of polarizability (Table 3). Other, computed ν$_{CO}$ Raman mode of vibration with high Raman intensity (118.94) and ν$_{NH}$ Raman mode (107.00) represent high polarizability. All the above-mentioned Raman active modes show the larger NLO activity in the I7C.

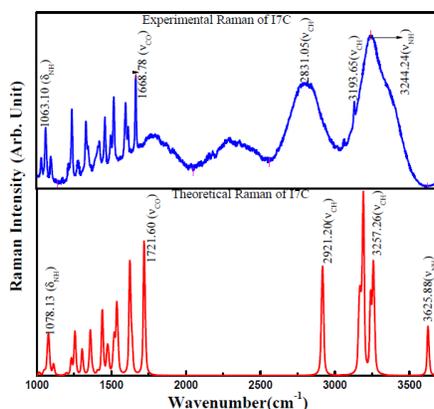

**Figure 3: Experimental and computed Raman spectra of I7C.**

*3.5. Chemical Reactivity*

In the present section, FMOs analysis is used illustrate the hardness/softness, chemical reactivity, and stability of probe molecule (Table 4). The HOMO shows the ability to donate an electron and LUMOs represent the ability to accept an electron. Koopmans' theorem has been used to calculate the different FMO parameters by using the following equation:

$$IP = -E_{HOMO} \quad (7)$$

$$EA = -E_{LUMO} \tag{8}$$

$$\eta = \frac{E_{LUMO} - E_{HOMO}}{2}, \quad S = \frac{1}{\eta} \tag{9}$$

$$CP = \frac{E_{HOMO} + E_{LUMO}}{2} \tag{10}$$

$$\omega = \frac{\mu^2}{2\eta} \tag{11}$$

$$\chi = \frac{(IP + EA)}{2} \tag{11}$$

The lower value of $\Delta\varepsilon$ (4.15 eV) for I7C than that of frequently used NLO active urea and KDP (Table 4) for current industry. This causes the CT in the molecule due to the formation of strong H-bonding between charged species. The NLO activeness also confirms by the higher value of ionization potential (*IP*), electronegativity ($\chi$), hardness ($\eta$), chemical potential (*CP*) and smaller softness (*S*). All the above mentioned values validate the strong candidature of I7C to be used as an NLO active material.

**Table 4: Chemical parameters of I7C.**

| S.No. | Molecular Properties | Values |
|---|---|---|
| 1. | HOMO | -6.284 |
| 2. | LUMO | -2.126 |
| 3. | Energy Gap ($\Delta\varepsilon$) | 4.158 |
| 4. | Ionization potential (*IP*) | 6.284 |
| 5. | Electron affinity (*EA*) | 2.126 |
| 6. | Chemical Potential (*CP*) | -4.205 |
| 7. | Electronegativity ($\chi$) | 4.205 |
| 8. | Softness (*S*) | 0.481 |
| 9. | Hardness ($\eta$) | 2.079 |

All values are in eV and S in (eV)$^{-1}$.

*3.6. UV-Vis Spectral Analysis*

To check the NLO properties of the I7C, experimental and computed absorption analysis have been performed and spectra are shown in Figure 4 and the corresponding computed transitions and *f* are given in the Table 5. The TPE-NH$_2$ system exhibit a broad absorption band in hydroxylic medium, one strong and wide band at 305-355 nm.

**Table 5: Experimental and computed electronic absorption table of I7C.**

| | | Theoretical | | | Experimental |
|---|---|---|---|---|---|
| | Transition | λ (nm) | E (eV) | (*f*) | λ (nm) |
| Absorption | S$_0$ → S$_1$ | 338.04 | 3.6677 | 0.0748 | 305-355 |
| | S$_0$ → S$_2$ | 325.77 | 3.805 | 0.0002 | |
| | S$_0$ → S$_3$ | 312.89 | 3.9625 | 0.0879 | |

Absorption wavelength λ (nm); excitation energies E (eV) and oscillator strengths (*f*)

Computed absorption bands for S$_0$→S$_1$ state is observed at 338.04 nm for S$_0$→S$_2$ state at 325.77 nm and for S$_0$→S$_3$ state at 312.89 nm with oscillator strength 0.0748, 0.0002 and 0.0879, respectively (Figure 4, Table 5). The *f* for S$_0$→S$_1$, S$_0$→S$_3$ transitions indicates their high possibilities of occurrence. These S$_0$→S$_1$ and S$_0$→S$_3$ transitions are assigned to π-π* and n-π* transitions, that makes the system unsaturated and enhances the possibility of active CT interactions. These transitions are responsible for the higher NLO activity in the molecule.

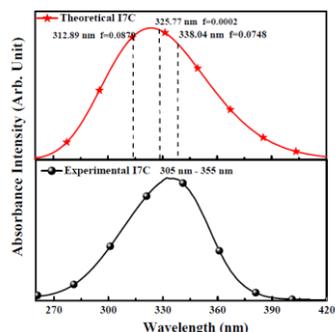

**Figure 4:** Experimental and computed electronic absorption spectra of I7C.

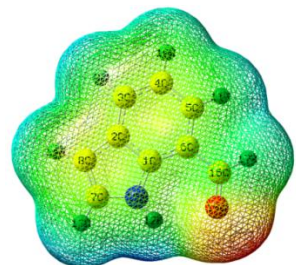

**Figure 5:** Molecular electrostatic potential of I7C molecule.

*3.8. Molecular Electrostatic Potential*

Molecular electrostatic potential (MEP) surface plot of electrostatic potential helpful in determining reactive site of the molecule [13]. "Maximum negative region with preferred site for electrophilic attack indicated by the blue color and the maximum positive region with preferred site for nucleophilic attack indicates as red color". A negative electrostatic potential over –NH group ($N_{15}$) represents the high possibility for the electrophilic reaction, whereas the red color surrounded C=O group indicates its repulsion (Figure 5). This behavior leads $I_{ra}CT$ from –NH group to C=O group and increases of the NLO activity of I7C.

**4. Conclusion**

The geometrical structure, charge variation, UV-Vis, Raman analysis of I7C have been carried out using DFT calculations. Computed polarizability, hyperpolarizability and active Raman modes have been used to predict the NLO properties of I7C. Variation in charge clearly indicates possibility of ICT interaction within the systems. The UV-Vis spectral analysis of I7C predicts the presence of $n \rightarrow \pi^*$ and $\pi \rightarrow \pi^*$ electronic transition and hence the variation of delocalized electron density within the molecule. Computed values of β is almost 6 times greater than the normally used reference material urea. MEP map and low energy gap also confirmed the NLO activity of I7C. The predicted results provide a better understanding into the NLO properties of I7C molecules to be used for many applications in optoelectronic and medical industries.